\documentclass[aps,nofootinbib,floatfix,amsmath,amssymb,superscriptaddress,preprint]{revtex4-1}

\usepackage{preamble}

\begin{document}

\title{Examining a right-handed quark mixing matrix with $b$-tags at the LHC}
\author{Andrew Fowlie}
\email{Andrew.Fowlie@KBFI.ee}
\affiliation{National Institute of Chemical Physics and Biophysics, Ravala 10,
Tallinn 10143, Estonia}

\author{Luca Marzola}
\email{Luca.Marzola@ut.ee}
\affiliation{Laboratory of Theoretical Physics, Institute of Physics, University of Tartu, Ravila 14c, Tartu 50411, Estonia}

\date{\today}


\begin{abstract}
Encouraged by a hint in a search for right-handed $W$ bosons at the LHC, we investigate
whether the unitarity of a right-handed quark mixing matrix and the equality of the left- and right-handed quark mixing
matrices could be tested at the LHC.
We propose a particular test, involving counting the numbers of $b$-tags in the final state, and
simulate the test at the event level with Monte-Carlo tools for the forthcoming \roots{13} LHC run.
We find that testing
unitarity with $20\invfb$ will be challenging; our test successfully rejects unitarity if the right-handed quark mixing matrix is
non-unitary, but only in particular cases.
On the other hand, our test may provide the first opportunity to test the unitarity
of a right-handed quark mixing matrix and with $3000\invfb$ severely constrains possible departures from unitarity in the latter.
We refine our previous work, testing the equality of quark mixing matrices, with full collider simulation.
With $20\invfb$, we are sensitive to mixing angles as small as $30\degree$, and with $3000\invfb$, angles as small as $7.5\degree$, confirming our preliminary analysis.
We briefly investigate testing the unitarity of the SM CKM matrix with a similar method by studying semileptonic $t\bar t$ production, concluding that systematics make it particularly difficult.
\end{abstract}

\maketitle

\section{Introduction}
Earlier this year, a small discrepancy in a search for right-handed (RH) $W$ bosons at the LHC\cite{Khachatryan:2014dka} lead to renewed
interest in left-right symmetric models\cite{Deppisch:2014zta,Jezo:2014wra,Aydemir:2014ama,Dekens:2014ina,Fowlie:2014mza,Senjanovic:2014pva,Dev:2014iva,Aguilar-Saavedra:2014ola,Heikinheimo:2014tba,Deppisch:2014qpa,Vasquez:2014mxa}. Left-right symmetric models were first discussed in the seventies\cite{Pati:1973uk,Pati:1974yy,Mohapatra:1974hk,Mohapatra:1974gc,Senjanovic:1975rk,Beg:1977ti,Senjanovic:1978ev} during the early years\cite{Georgi:1974sy,Fritzsch:1974nn,Gursey:1975ki,Buras:1977yy} of grand unification theory (GUT). At moderate energies, such models are described by the gauge symmetries
\beq
\text{SU}(2)_L \times \text{SU}(2)_R,
\eeq
and a discrete symmetry, such as parity or charge conjugation, playing the role of left-right symmetry, all of which are spontaneously broken at low energy. These symmetries require RH analogues of the Standard Model (SM) $W$  and $Z$ bosons and of the SM neutrinos. In the SM, the flavor structure of left-handed (LH) quark charged interactions is governed by the CKM matrix\cite{Kobayashi:1973fv,Cabibbo:1963yz}; similarly, in a left-right symmetric model, distinct quark mixing matrices describe the flavor structure of LH and RH quark charged interactions\cite{Zhang:2007fn,Zhang:2007da}. In our previous work\cite{Fowlie:2014mza}, we found that future experiments at the LHC could detect discrepancies between the LH and RH quark mixing matrices. In this work, we refine our previous analysis of left-right symmetry with collider simulations and turn our attention to the unitarity of the RH quark mixing matrix. The unitarity of a three-by-three matrix,
\beq
V V^\dagger = \id,
\eeq
implies six orthogonality constraints (widely represented by six unitarity triangles) and three normalization constraints. A common strategy for testing the unitarity of the SM CKM matrix \see{and:2013lpa,CKM,Battaglia:2003in} is to make separate, precise measurements of CKM matrix elements. For example, the up-type matrix elements, $V_{uq}$, can be independently extracted from rare decays. The measurements are in excellent agreement with a unitarity normalization constraint\cite{CKM_u},
\beq
|V_{ud}|^2+|V_{us}|^2+|V_{ub}|^2 = 0.9999 \pm 0.0006.
\eeq
The normalization constraint predicts that this quantity is exactly one, while the presence of new physics, especially a fourth generation of quarks, might cause a departure from unitarity in the three-by-three CKM matrix.

In this paper we, however, suggest a different strategy for testing the unitarity of a RH quark mixing matrix, if a RH sector is discovered in the future. Because a RH $W$ boson (henceforth \rw boson) would have to be heavier than about $2\tev$ to have escaped direct detection\cite{W_R}, or about $2\tev$ to explain the anomaly in \refcite{Khachatryan:2014dka}, decays to all SM quarks are kinematically allowed, and even the top quark can be regarded as approximately massless, $m_t/M_{\rw} \lesssim 0.1$. At a proton collider such as the LHC, jets from the decay of a \rw boson would be harder than typical SM backgrounds. Thus, we propose that the unitarity of the RH quark mixing matrix could be tested by analyzing a \rw boson's branching fractions at the LHC. In light of \refcite{Mangano:2014xta}, in which the authors stress the enormous numbers of $W$ bosons expected to be produced at the LHC, we briefly discuss a similar strategy for the SM CKM matrix in \refsec{Sec:SM_W}.

To parameterize a non-unitary RH quark mixing matrix \see{Fritzsch:1986gv}, we first write a product of rotations on six planes in the basis of the quark fields $(d,s,b,f)^T$, where $f$ is a very heavy, undiscovered fourth-generation down-type quark \see{Cetin:2011aa}:
\beq
\label{Eq:V4x4}
V = R({\theta^R_{34}}) \times R({\theta^R_{24}}) \times R({\theta^R_{23}}) \times R({\theta^R_{14}}) \times R({\theta^R_{13}}) \times R({\theta^R_{12}}).
\eeq
This is a unitary four-by-four RH quark mixing matrix. We introduce a heavy fourth-generation quark only to parameterize our matrix; we do not assume that it exists in Nature. The non-unitary three-by-three RH quark mixing matrix is found by omitting the row and column corresponding to the fourth-generation quark,
\beq
V_R = \[V\]_{3\times3} \tand V_R^{\phantom{\dagger}} V_R^\dagger \neq \id.
\eeq
All elements of a non-unitary quark mixing matrix parameterized in this manner, $V_R$, and all elements of $V_R^{\phantom{\dagger}} V_R^\dagger$ are less than or equal to one in absolute value. We parameterize a non-unitary quark mixing matrix in this restrictive manner because a fourth-generation quark is a realistic scenario in which the full quark mixing matrix is unitary. An arbitrary non-unitary quark mixing matrix corresponds to non-conservation of probability. To give concrete examples of our methodology, for the sake of simplicity, we often assume either that there is an angle governing the mixing between the first three generations and an angle governing the mixing with the fourth generation:
\beq\label{Eq:2Angle}
\theta_3 \equiv \theta^R_{12} = \theta^R_{13} = \theta^R_{23} \tand \theta_{4} \equiv \theta^R_{14} = \theta^R_{24} = \theta^R_{34},
\eeq
or that the mixing angles in \refeq{Eq:V4x4} are equal to a universal mixing angle:
\beq\label{Eq:1Angle}
\theta \equiv \theta_3 \equiv \theta_4.
\eeq

\section{Methodology}
We consider \rw bosons produced at the LHC at \roots{13} and decaying via the chain
\beq
\label{Eq:DecayChain}
pp \to \rw \to e \nu_e^R  \to e e \rw^* \to e e j j.
\eeq
The details of our simulations of such a process are forthcoming. The final \rw boson is off-shell (indicated by an asterisk). The jets associated with the $W^*_R$ decay could carry zero, one or two $b$-tags, resulting in three categories, \ie three separate counting experiments. Because we assume that the RH electron neutrino is the lightest RH neutrino, the off-shell \rw boson in this decay chain cannot decay leptonically; the hadronic branching fraction for the off-shell \rw boson is $100\%$. If the RH quark mixing matrix is unitary, the relevant branching fractions and cross sections are described by two parameters: the top-bottom RH quark mixing matrix element,
\beq
|V^R_{tb}|^2 = \cos^2\theta^R_{13} \cos^2\theta^R_{23},
\eeq
and the cross section, which is a function of the \rw boson's mass, the RH neutrino masses, the RH gauge coupling, and the RH quark mixing matrix,
\beq
\sigma(pp \to \rw \to e e j j ) = f(g_R, M_{\rw}, m_{\nu_R}, V_R),
\eeq
and well-known SM parameters. The cross section determines the total number of expected \rw boson hadronic decays, whereas the top-bottom RH mixing matrix element, $|V^R_{tb}|^2$, determines the fraction of hadronic decays with zero, one or two $b$-tags\cite{Fowlie:2014mza}. Thus, there are three independent measurements described by only two parameters and the RH quark mixing matrix may be ``over-fitted,'' implying that we may test its unitarity.

Let us elaborate upon this claim. Neglecting quark masses, the \rw boson's branching fractions into two electrons and first- and second-generation quarks, $q$, or third-generation quarks, $t$ and $b$, are
\begin{align}
\text{BR}(\rw\to e e t b) &\propto \tfrac13 |V^R_{tb}|^2\\
\text{BR}(\rw\to e e q q)  &\propto \tfrac13 |V_{ud}|^2+\tfrac13|V_{us}|^2 +\tfrac13 |V_{cd}|^2+\tfrac13|V_{cs}|^2 = \tfrac13(1+|V^R_{tb}|^2)\\
\text{BR}(\rw\to e e q t/b) &\propto \tfrac13|V_{ub}|^2+\tfrac13|V_{cb}|^2+\tfrac13|V_{dt}|^2+\tfrac13|V_{st}|^2=\tfrac23(1-|V^R_{tb}|^2)
\end{align}
The choices of $g_R$, $M_{\rw}$, $m_{\nu_R}$, and free parameters in $V_R$ other than $|V^R_{tb}|$ can affect the total number of hadronic \rw decays, but cannot affect the relevant branching fractions or the expected numbers of events with zero, one or two $b$-tags. By fitting those parameters, one can tune the total number of observed events from the chain in \refeq{Eq:DecayChain}, and by fitting $|V^R_{tb}|^2$, one can tune, say, the ratio of one and two $b$-tag events. A remaining independent quantity, the number of zero $b$-tag events, cannot be tuned. Thus, a unitary quark mixing matrix is ``over-fitted,'' and we can check its unitarity through this undetermined quantity. In fact, unitarity results in restrictions in the branching fractions:
\beq\label{Eq:Restriction}
\frac{\text{BR}(\rw\to e e q q)}{\text{BR}(\rw\to e e q t/b)} \geq \frac12 \tand \frac{\text{BR}(\rw\to e e q q)}{\text{BR}(\rw\to e e tb)} \geq 2.
\eeq
Our methodology, however, will test forbidden correlations among branching fractions, as well as these trivial restrictions.

If we assume that the RH mixing matrix is equal to the LH CKM matrix, the total number of events from the decay chain in \refeq{Eq:DecayChain} is governed by a free parameter (a cross section), but the $b$-tag distribution is determined by the LH CKM mixing angles. With three measurements and a single free parameter, as in our previous analysis\cite{Fowlie:2014mza}, we may test the equality of the LH and RH quark mixing matrices.

\subsection{Collider simulation of right-handed $W$ boson signal}\label{Sec:CS}
We simulated \rw boson production and subsequent decay via \refeq{Eq:DecayChain} at the LHC at \roots{13} with Monte-Carlo tools. With \texttt{MadGraph-2.2.1}\cite{Alwall:2014hca}, we calculated the relevant matrix elements for a simple left-right symmetric \texttt{FeynRules}\cite{Alloul:2013bka} model, based upon well-tested \texttt{FeynRules} models, with Lagrangian:
\beq
\mathcal{L} = \mathcal{L}_{\text{SM}} - \frac{g_R}{\sqrt{2}}  \bar\nu_{\ell_{iR}} \gamma_\mu W_R^\mu \ell_{iR} - \frac{g_R}{\sqrt{2}}  V^R_{ij} \bar u_R^i \gamma_\mu W_R^\mu d_R^j + M_{W_R}^2 |W_R|^2 -\frac{m_{\nu_{\ell_{iR}}}}{2} \bar\nu_{\ell_{iR}}^c \nu_{\ell_{iR}}^{\phantom{c}} + \text{h.c.}
\eeq
Our analysis is not sensitive to a lepton mixing matrix (absent in our Lagrangian). The presence of a lepton mixing matrix could suppress our decay chain, reducing the total number of expected events. The latter could, however, be compensated by increasing the RH coupling.

With those matrix elements, we generated events with \texttt{Pythia}\cite{Sjostrand:2007gs,Sjostrand:2006za}, linked with the \texttt{PGS4}\cite{PGS} detector simulator with the CMS detector input card and the anti-$k_T$ clustering algorithm\cite{Cacciari:2008gp} with a distance parameter of $r=0.5$. The result was four sets of $10\,000$ events, composed of reconstructed objects (jets, electrons, muons \etc) corresponding to four distinct \rw boson decay channels in \refeq{Eq:DecayChain}, in which the off-shell \rw bosons decay to:%
\begin{inparaenum}[(1)]
\item two first- or second-generation quarks (light quarks), denoted $qq$,
\item a light quark and a bottom quark ($qb$),
\item a light quark and a top quark ($qt$), and
\item a top and a bottom quark ($tb$).
\end{inparaenum}

Upon those events, we imposed the following selections on the jets and electrons, based upon those in the CMS search in \refcite{Khachatryan:2014dka}:
\begin{description}
\item[Electrons]\mbox{}\\[-1.5\baselineskip]
\begin{enumerate}
\item We vetoed electrons with pseudo-rapidity $\eta>2.5$.
\item We required exactly two electrons (of any charge).
\item We required that the hardest electron had transverse momentum $P_T>60\gev$ and the second hardest electron had $P_T>40\gev$.
\item We required that the invariant mass of the two hardest electrons was $M_{ee} > 200\gev$.
\end{enumerate}
\item[Jets]\mbox{}\\[-1.5\baselineskip]
\begin{enumerate}
\item We vetoed jets with $\eta>2.5$.
\item We required at least two jets with $P_T>40\gev$, and picked the hardest two.
\item We required that the invariant mass of the two hardest electrons and the two hardest jets was $M_{eejj} > 600\gev$.
\item We $b$-tagged $b$-jets with a probability of $\epsilon=0.7$.
\item We $b$-tagged other jets with a probability of $\rho=0.01$.\footnote{Although \texttt{PGS4} permits detailed, momentum dependent tagging algorithms, all such algorithms require that \texttt{PGS4} is an ``oracle'' that reveals a jet's true flavor and calculate a probability that a jet is tagged. We picked a crude algorithm to simplify our analysis, but which is a reasonable approximation to an experimental analysis.}
\end{enumerate}
\end{description}
By counting the numbers of events that passed our selections, we estimated the selection efficiencies for the \rw boson signal.\footnote{We scrutinized events from \texttt{PGS4} in the \texttt{LHCO} format with a new code, \texttt{LHCO\_reader}\cite{LHCO_reader}.}

We picked masses and couplings for the RH sector such that the \rw boson could explain the small excess observed in events with two electrons and two jets with an invariant mass of about $2\tev$\cite{Khachatryan:2014dka,Heikinheimo:2014tba,Deppisch:2014qpa,Aguilar-Saavedra:2014ola}. In particular, we decoupled tau and muon RH neutrino masses, but set the electron RH neutrino mass, $m_{\nu_R^e} = \tfrac12 M_{\rw}$, such that \rw bosons decayed via electron-neutrinos. We imposed $M_{\rw}=2\tev$, $g_R = \tfrac12 g_L$ and a diagonal RH quark mixing matrix, $V_R=\id$. We find that $\sigma(pp \to \rw \to eejj) \simeq 13.9\,\text{fb}$ at \roots{13}. At \roots{8}, we find that $\sigma(pp \to \rw \to eejj) \simeq 2.2\,\text{fb}$, below the experimental upper limit of $2.29\,\text{fb}$\cite{Khachatryan:2014dka}.

The Monte-Carlo simulations are summarized by a four-by-three matrix of efficiencies (including $b$-tagging, selection, and detector efficiencies), corresponding to our four decay channels and three $b$-tag categories. The matrix relates the numbers of \rw bosons decaying in a particular manner with the number of selected events in each $b$-tag category:
\beq\label{Eq:Eff_Matrix}
\begin{pmatrix}
\text{Number $0$ $b$-tags} \\
\text{Number $1$ $b$-tags} \\
\text{Number $2$ $b$-tags} \\
\end{pmatrix}
=
\begin{pmatrix}
0.59&  0.24&  0.18&  0.08\\
0.02&  0.32&  0.12&  0.14\\
0.00&  0.01&  0.01&  0.06\\
\end{pmatrix}
\begin{pmatrix}
\text{Number $\rw\to eeqq$} \\
\text{Number $\rw\to eeqb$} \\
\text{Number $\rw\to eeqt$} \\
\text{Number $\rw\to eetb$} \\
\end{pmatrix}.
\eeq
If our selections and $b$-tagging algorithms resulted in no impurities, the matrix would read:
\beq
\begin{pmatrix}
a &  0 &  0 & 0 \\
0 &  b &  c & 0 \\
0 &  0 &  0 & d \\
\end{pmatrix}.
\eeq
If our efficiencies were $100\%$, the constants in this matrix would be equal to one. Our selection efficiency for $V_R = V_L$ without a cut on the invariant mass $M_{eejj}$ is about $75\%$, which compares reasonably with $78\%$ quoted in \refcite{Khachatryan:2014dka}.

We select the two hardest jets, which we assume originate from the \rw boson. There is, however, an appreciable chance that our procedure selects a rogue jet, damaging our purities. A \rw boson decaying to a top quark is especially problematic, because the top itself decays, $t\to Wb$. If the $W$ boson decays leptonically, the event might be vetoed as it could contain more than two electrons. Even if the $W$ decays hadronically, because the $W$ boson carries away momentum, the invariant mass of the \rw boson might not be reconstructed. For example, the selection efficiencies for the $\rw\to eetb$ decay sum to about $30\%$ of which about $80\%$ are incorrectly categorized (see the fourth column in the matrix in \refeq{Eq:Eff_Matrix}). In principle, it might be possible to improve our efficiencies with a dedicated analysis.

\subsection{Standard Model backgrounds}
The dominant SM backgrounds are dileptonic $t \bar{t}$ and Drell-Yan, which potentially contaminate each of our $b$-tags categories due to the imperfections in the $b$-tagging algorithm. These processes have substantial cross sections and impair our \rw search in the tails of their kinematic distributions, yielding technical difficulties in dedicated simulations. Instead of performing a specific analysis we then choose to model our backgrounds as follows. We first consider a number of background events that after all selections, efficiencies, etc, matches the data-driven prediction in \refcite{Khachatryan:2014dka} at \roots{8}, where the background is estimated from a sideband.
In order to account for the different center-of-mass energy considered in our study, we then apply to the quoted number of background events a further rescaling factor given by the ratio or the relevant cross sections calculated at 8 and 13 TeV. In order to give a conservative estimate, we furthermore assumed that all background events result in a $b$-tag distribution identical to that of $t \bar{t}$ production.
Finally, as a crosscheck, we increased the expected background by a factor of $50\%$, finding that our conclusions are robust and the sensitivity of our test weakens only by a few degrees.


\subsection{Statistical treatment}
We examine the RH quark mixing matrix with a methodology similar to that in our previous work\cite{Fowlie:2014mza}, with which we tested the equality of the LH and RH quark mixing matrices. The statistical treatment is based upon Poisson statistics of three independent counting experiments: the numbers of hadronic \rw boson decays resulting in zero, one or two $b$-tags. We conduct two similar statistical tests: a test of the equality of the LH and RH quark mixing matrices, and a test of the unitarity of the RH quark mixing matrix.

\subsubsection{Testing the unitarity of the right-handed quark mixing matrix}
We consider two hypotheses in our statistical test of the unitarity of the RH quark mixing matrix:
\begin{itemize}
\item The null hypothesis, $H_0$: the RH quark mixing matrix is unitary, that is,
\beq
V_R^{\phantom{\dagger}} V_R^\dagger = \id.
\eeq
In this case, the numbers of expected events in each $b$-tag category resulting from \refeq{Eq:DecayChain} are described by a cross section, $\sigma$, and the top-bottom quark mixing matrix element, $|V^R_{tb}|^2$.
\item The alternative hypothesis, $H_1$: the RH quark mixing matrix is non-unitary. In this case, the numbers of expected events are described by a cross section, $\sigma$, and the independent elements of the non-unitary RH quark mixing matrix.
\end{itemize}
To test whether future LHC experiments could reject the null hypothesis that the RH quark mixing matrix is unitary, we construct a log-likelihood ratio (LLR) test statistic and find $p$-values, assuming that Nature is described by a non-unitary RH quark mixing matrix.

Having picked a particular non-unitary matrix, we:
\begin{enumerate}
\item Calculate the expected numbers of signal and background events in the zero, one and two $b$-tag categories, $\lambda_i$, with $i=\text{$0$, $1$ or $2$}$. For simplicity, we assume that the production cross section is independent of the quark mixing matrix and always equal to that obtained in \refsec{Sec:CS}. This is because changes in the production cross section due to the quark mixing matrix can be compensated by altering the RH coupling.
\item Simulate a counting experiment, by drawing $1000$ samples, $\{o_i\}$, from Poisson distributions with mean $\lambda_i$, $o_i\sim\text{Po}(\lambda_i)$.
\item For each sample, we calculate our LLR for our hypotheses:
\beq
\text{LLR} = -2\ln \frac{\max \mathcal{L}(\{o_i\}\,|\,H_0, \overbar\sigma,\overbar{|V^R_{tb}|^2})}{\max\mathcal{L}(\{o_i\}\,|\,H_1,\overbar\sigma,\overbar{V_R})},
\eeq
where the likelihood functions, $\mathcal{L}$, are Poisson distributions and a line above a parameter indicates that the value of that parameter is chosen such that the likelihood is maximized. In the numerator, $V_{tb}^R$ refers to a single element of the quark mixing matrix, whereas in the denominator, $V_R$ is the RH quark mixing matrix.

We found the LLR distribution from MC. The $p$-value is the probability of obtaining such a large LLR by chance, were the null hypothesis true --- the area under the right-hand-side tail in the LLR distribution.

\item Lastly, we calculate the median and $68\%$ confidence interval for the $p$-value, by considering all of our pseudo-experiments.
\end{enumerate}

\subsubsection{Testing the equality of left- and right-handed quark mixing matrices}
Similarly, we consider two hypotheses in our statistical test of the equality of LH and RH quark mixing matrices:
\begin{itemize}
\item The null hypothesis, $H_0$: the RH quark mixing matrix is equal to the LH quark mixing matrix, that is,
\beq
V_R  = V_L
\eeq
In this case, the numbers of expected events in each $b$-tag category resulting from \refeq{Eq:DecayChain} are described by a cross section, $\sigma$. The fractions of events in each $b$-tag category are determined by the (fixed) LH quark mixing matrix.

\item The alternative hypothesis, $H_1$: the RH quark mixing matrix is an arbitrary, non-unitary three-by-three matrix. In this case, the numbers of expected events are described by a cross section, $\sigma$, and the independent elements of the non-unitary RH quark mixing matrix.
\end{itemize}

Assuming that Nature is described by a unitary RH quark mixing matrix, we follow an identical sequence of steps as before, though with a different LLR test statistic:
\beq
\text{LLR} = -2\ln \frac{\max \mathcal{L}(\{o_i\}\,|\,H_0, \overbar\sigma)}{\max\mathcal{L}(\{o_i\}\,|\,H_1,\overbar\sigma,\overbar{V_R})}.
\eeq
As previously, the likelihood functions, $\mathcal{L}$, are Poisson distributions, and a line above a parameter indicates that the value of that parameter is chosen such that the likelihood is maximized.

\section{Unitarity of Standard Model CKM matrix}\label{Sec:SM_W}
It is possible to extract the SM CKM matrix element $V_{tb}$ from single top production\cite{Khachatryan:2014iya} and, if one assumes unitarity, from $t\bar t$ production\cite{Khachatryan:2014nda}. We believe that it is, in principle, possible to test the unitarity of the SM CKM matrix and constrain CKM matrix elements at the LHC by analyzing $W$ boson decays. The semileptonic $t\bar t$ process,
\beq
pp\to t \bar t \to b \bar b W W \to b\bar b \ell \nu q q^\prime,
\eeq
could be reconstructed with two $b$-tags, a lepton and a top-tagging algorithm. If the CKM matrix is unitary, the total number of $W\to qb$ and $W\to qq$ events ought to follow
\beq
n(\text{$qb$ events}) = \text{const.} \times |V_{tb}|^4 (1-|V_{tb}|^2) \tand n(\text{$qq$ events}) = \text{const.} \times |V_{tb}|^4 (1+|V_{tb}|^2),
\eeq
whereas if the CKM matrix is not unitary,
\beq
n(\text{$qb$ events}) = \text{const.} \times  R^2 \eta \tand n(\text{$qq$ events}) = \text{const.} \times  R^2 (1 - \eta),
\eeq
where we define $R \equiv |V_{tb}|^2 / (|V_{ts}|^2 + |V_{td}|^2 + |V_{tb}|^2)$ and $\eta \equiv (|V_{ub}|^2 + |V_{cb}|^2) / (|V_{ub}|^2 + |V_{cb}|^2 + |V_{ud}|^2 + |V_{us}|^2 + |V_{us}|^2 + |V_{cs}|^2)$. These decays are distinguishable in an experiment with a $b$-tagging algorithm (because $W\to tb$ is kinematically forbidden, there is no two $b$-tag category). With two measurements, the $V_{tb}$ element is over-constrained. We find, however, that regardless of large statistics of about $10^9$ $W$ bosons, the test is thwarted by systematic errors in the $t\bar t$ cross section and backgrounds. It may, however, be possible to extract and verify information about the CKM matrix in such an analysis.

\section{Results}
Let us recapitulate our goal.  Assuming that a \rw boson is discovered in the future, we want to know whether a \roots{13} LHC analysis of \rw bosons is sensitive to the unitarity of the RH quark mixing matrix or the equality of the LH and RH quark mixing matrices. We consider the cases separately, in the  following two sections. We initially consider $20\invfb$ of data at \roots{13}, before investigating the ultimate statistical power of our tests with $3000\invfb$, although their sensitivities may improve if efficiencies and purities, especially in the two $b$-tag category, are refined.

\subsection{Unitarity}\label{sec:unitarity}
\begin{figure}[t]
\centering
\includegraphics[width=.48\textwidth,valign=t]{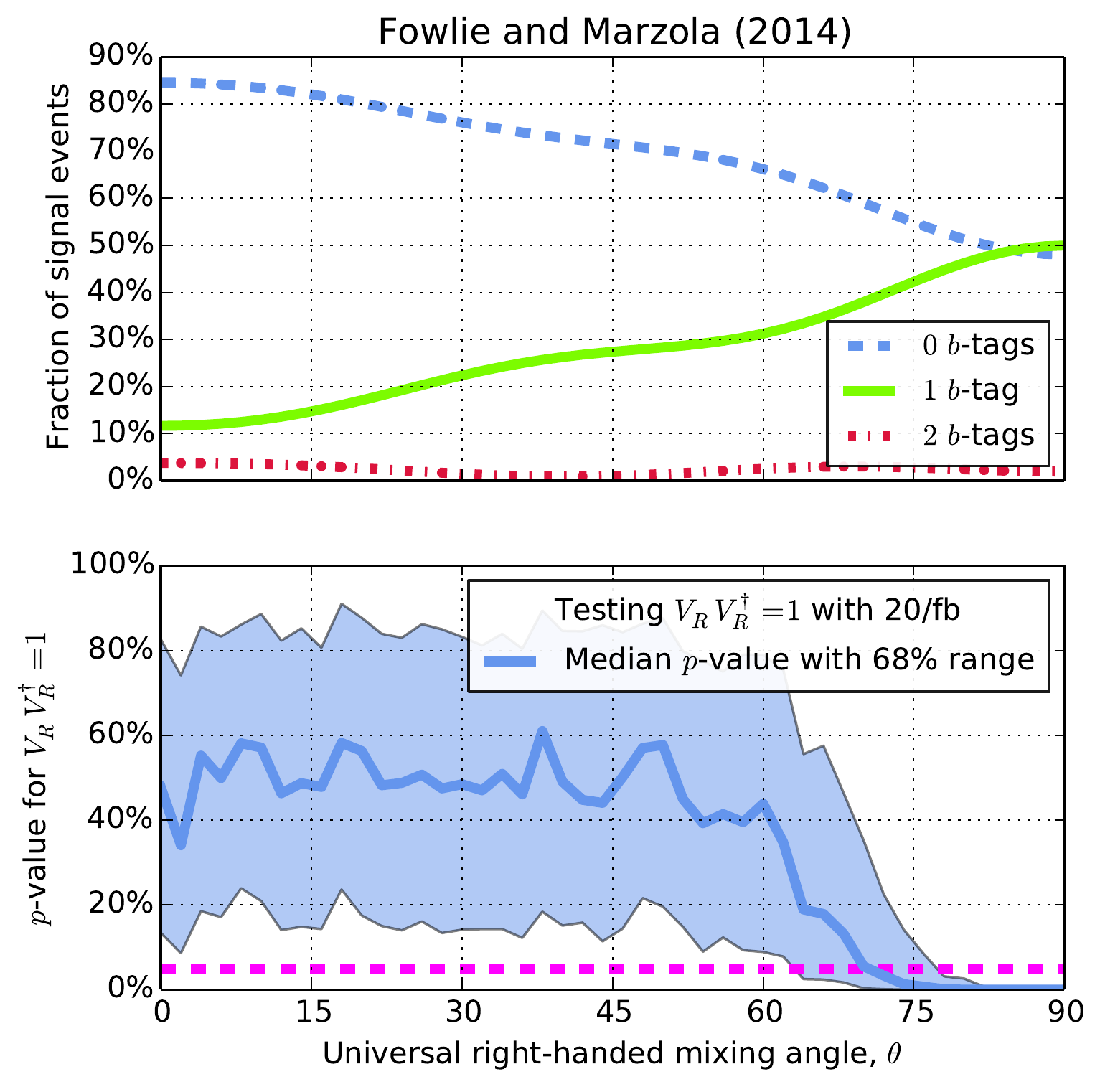}
\caption{Potential to exclude a unitary RH quark mixing matrix, $V_R^{\protect\phantom{\dagger}} V_R^\dagger = \id$, as a function of a universal mixing angle by counting $b$-tags at the LHC with \roots{13}, $\int\mathcal{L}\sim20\invfb$. When the $p$-value falls below $5\%$, we can reject the null hypothesis, that the RH quark mixing matrix is unitary, with $95\%$ confidence. The upper panel shows the expected fractions of $b$-tags in the non-unitary case.
}
\label{fig:UNIT20}
\end{figure}

We plot the potential to exclude $V_R^{\phantom{\dagger}} V_R^\dagger = \id$ against a single universal mixing angle, $\theta\equiv\theta_3\equiv\theta_4$, with $20\invfb$ in \reffig{fig:UNIT20}. Whereas the upper panel shows the expected fraction of signal events in each $b$-tag category, in the lower panel we plot the median $p$-value (solid blue line), as well as its $68\%$ range (filled blue band). If the former drops below $5\%$ (dashed magenta line), in the majority of cases, we could reject the null hypothesis that $V_R = V_L$ with at least $95\%$ confidence. As we can see, if the mixing angle is greater than about $75\degree$, a unitary matrix cannot reproduce the fractions of $b$-tags, because \refeq{Eq:Restriction} is violated. If the angle is less than about $75\degree$, however, it is possible to find a unitary matrix that perfectly mimics the non-unitary matrix (though the latter is not necessarily approximately unitary), and the hypotheses cannot be distinguished. Our test's poor sensitivity results from the low selection efficiencies for the two $b$-tag category in \refeq{Eq:Eff_Matrix}, which result in a $s/\sqrt{b}$ ratio in that category of about one. With such poor sensitivity in that category, there are effectively only two $b$-tag categories: zero and one $b$-tags. The numbers of events in two categories can be perfectly fitted in the null hypothesis (a unitary matrix), with its two free parameters. In fact, even if all mixing angles with the fourth generation are maximal, $\theta_4 = 90\degree$, unitarity can only be rejected if the ordinary mixing angles are greater than about $45\degree$.

\begin{figure}[t]
\centering
\includegraphics[width=.48\textwidth,valign=t]{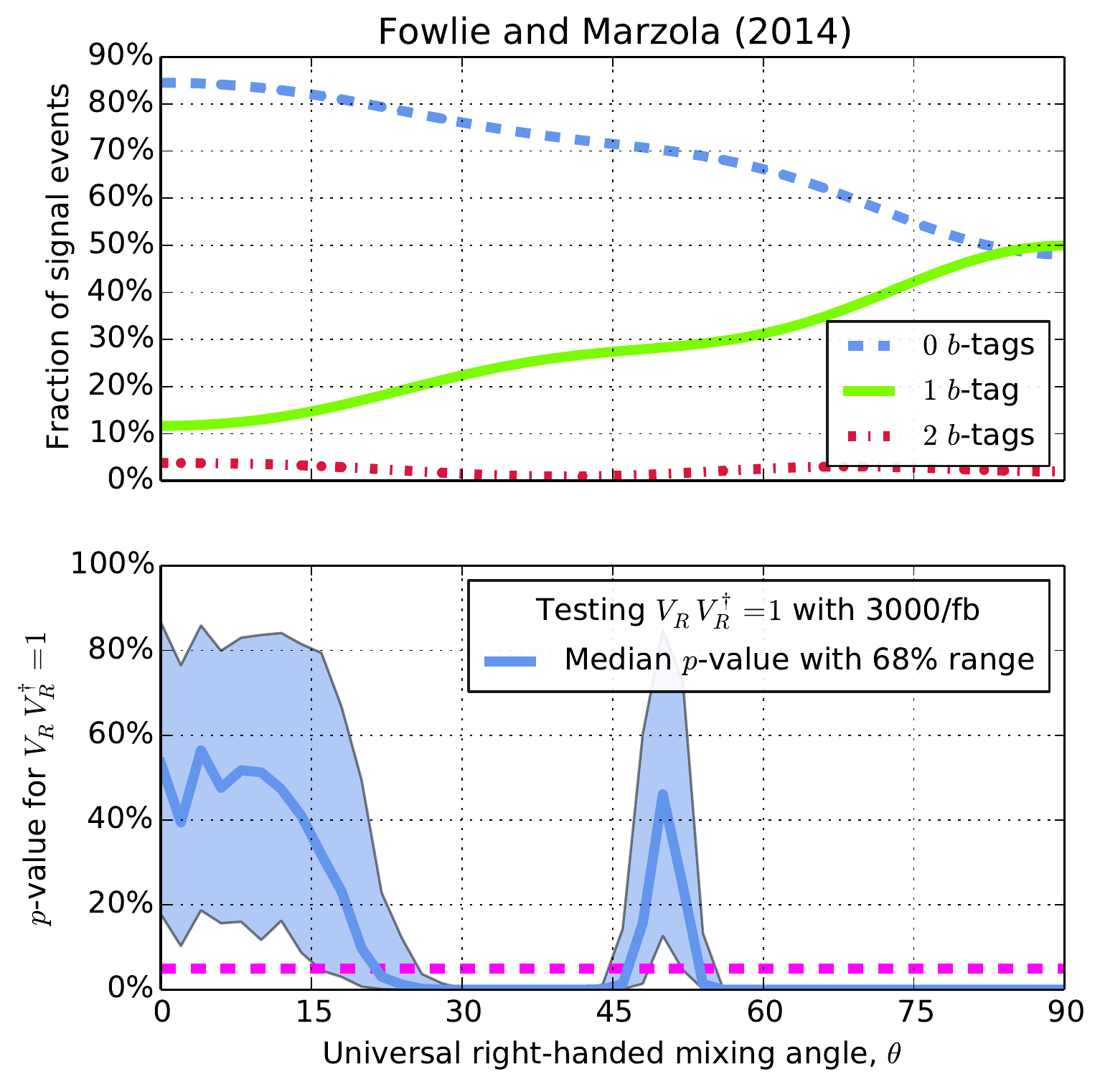}
\caption{As in \reffig{fig:UNIT20}, but with $3000\invfb$. Potential to exclude a unitary RH quark mixing matrix, $V_R^{\protect\phantom{\dagger}} V_R^\dagger = \id$, as a function of a universal mixing angle by counting $b$-tags at the LHC with \roots{13}, $\int\mathcal{L}\sim3000\invfb$. When the $p$-value falls below $5\%$, we can reject the null hypothesis, that the RH quark mixing matrix is unitary, with $95\%$ confidence. The upper panel shows the expected fractions of $b$-tags in the non-unitary case.
}
\label{fig:UNIT3000}
\end{figure}

Having considered $20\invfb$, we now analyze a scenario with $3000\invfb$ of data, corresponding to the entire LHC operation. With such large statistics, the two $b$-tag category finally helps to split the hypotheses, as we show in \reffig{fig:UNIT3000}, in which we plot \reffig{fig:UNIT20} but with $3000\invfb$ rather than $20\invfb$. As well as the trivial exclusion of unitarity resulting from \refeq{Eq:Restriction} above about $50\degree$, a region at between about $20\degree$ and $45\degree$ is excluded, as ratios between the various three categories of $b$-tags cannot be achieved with a unitary matrix. The narrow region around $\theta\simeq50\degree$ is not excluded because of the interplay between the RH mixing matrix and the RH coupling constant. This degeneracy could be broken if the latter were measured from a RH $Z$ boson resonance (and the RH lepton mixing matrix was assumed or known). As a remark, we parameterized a non-unitary quark mixing matrix as a sub-block of a four-by-four unitary matrix. If we generated RH quark mixing matrices with random entries from zero to one that satisfy the restrictions in \refeq{Eq:Restriction}, with $3000\invfb$, we could reject unitarity in about $75\%$ of cases.

\begin{figure}[t]
\centering
\includegraphics[width=.48\textwidth,valign=t]{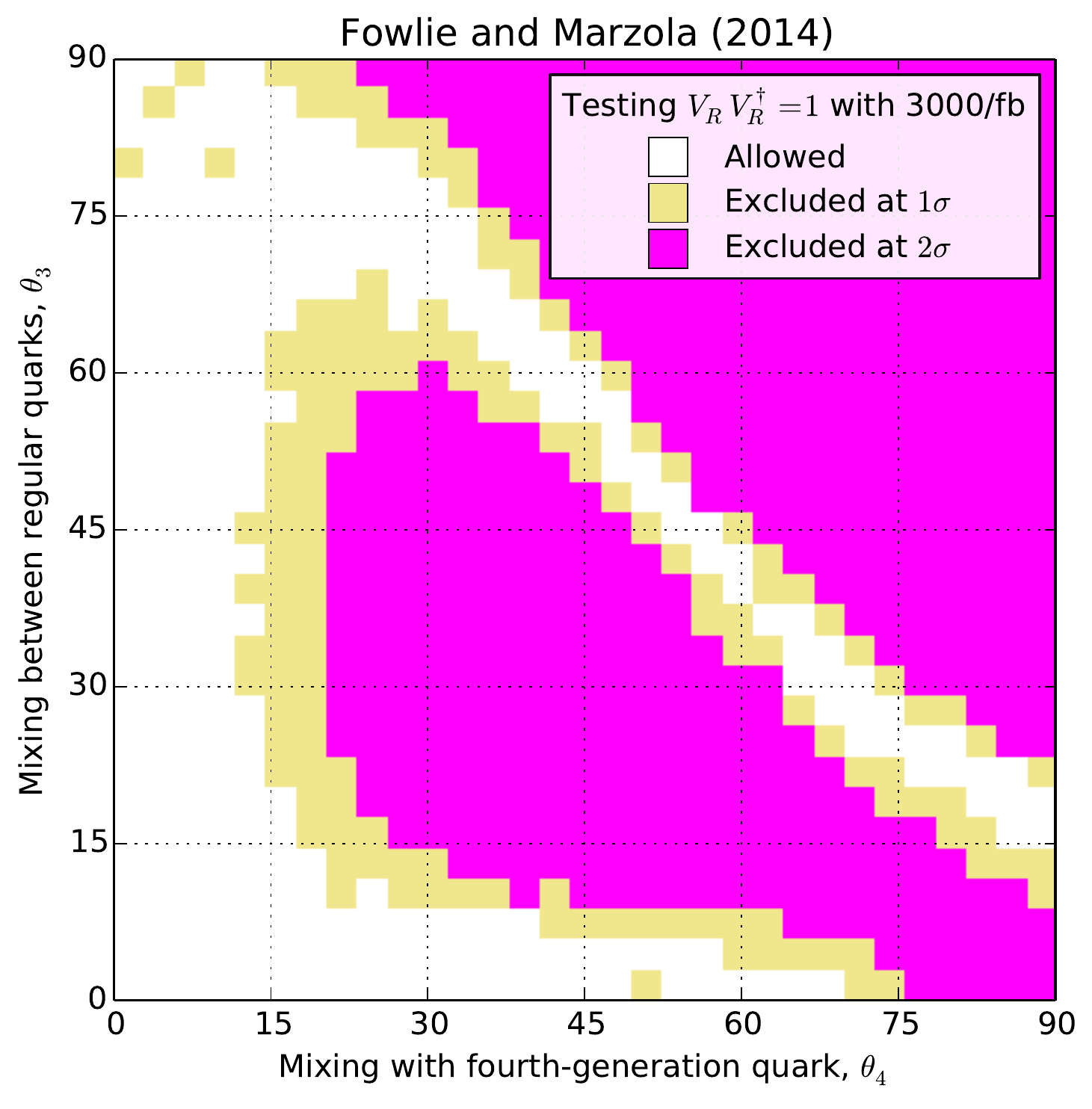}
\caption{Potential to exclude a unitary RH quark mixing matrix, $V_R^{\protect\phantom{\dagger}} V_R^\dagger = \id$, on the $(\theta_3,\theta_4)$ plane from counting $b$-tags at the LHC with \roots{13}, $\int\mathcal{L}\sim3000\invfb$. For example, if the median $p$-value is below $5\%$, we plot $2\sigma$ exclusion. The angle $\theta_4$ is a universal mixing angle with a fourth generation quark, whilst $\theta_3$ is a universal mixing angle among the regular three generations of quark.
}
\label{fig:2D_UNIT3000}
\end{figure}

Our results for unitarity are sensitive to the mixing angles between the first three generations of RH quarks. In \reffig{fig:2D_UNIT3000}, we plot the median exclusion on the $(\theta_3,\theta_4)$ plane with $3000\invfb$. The angle $\theta_{4}$ parameterizes the departure from unitarity (if $\theta_{4}=0$, the RH quark mixing matrix is unitary) and the angle $\theta_3$ describes the mixing among the SM RH quarks (see \refeq{Eq:2Angle}). We see from \reffig{fig:2D_UNIT3000} that to reject $V_R^{\phantom{\dagger}} V_R^\dagger = \id$ at $95\%$ confidence, we require that the departure from unitarity is substantial, $\theta_4\gtrsim75\degree$ or that other mixing angles are moderate, $\theta_3 \gtrsim 15\degree$. The latter possibility is encouraging; if the RH mixing angles among the three generations of quark are slightly greater than the Cabbibo angle in the LH mixing matrix (about $15\degree$), unitarity could be rejected if the mixing angle with a fourth-generation quark is greater than about $25\degree$.

In summary, testing the unitarity of a RH quark mixing matrix is challenging with $20\invfb$ due to the poor efficiencies for the two $b$-tag category and moderate backgrounds. On the other hand, our method could cast powerful bounds on unitarity with $3000\invfb$ and if future experiments confirmed the anomaly hinting at a \rw boson, it could aid the interpretation of the latter.

\subsection{Equality of left- and right-handed quark mixing matrices}
In a similar fashion to that in \refsec{sec:unitarity}, we present results for our test of the equality of the LH and RH quark mixing matrices at \roots{13} in two scenarios: a limited run of $20\invfb$ of data and a high-luminosity run of $3000\invfb$, to investigate the immediate and ultimate power of our test. In our previous work in \refcite{Fowlie:2014mza}, which we refer to as FM1, as a preliminary to this work we demonstrated this test with an approximate analysis. We return to it now with a complete MC collider simulation. In this section, we vary only the mixing angle between ordinary RH quarks ($\theta_3$) and fix the mixing with a heavy fourth-generation quark to zero ($\theta_4 = 0$).

\begin{figure}[t]
\centering
\includegraphics[width=.48\textwidth,valign=t]{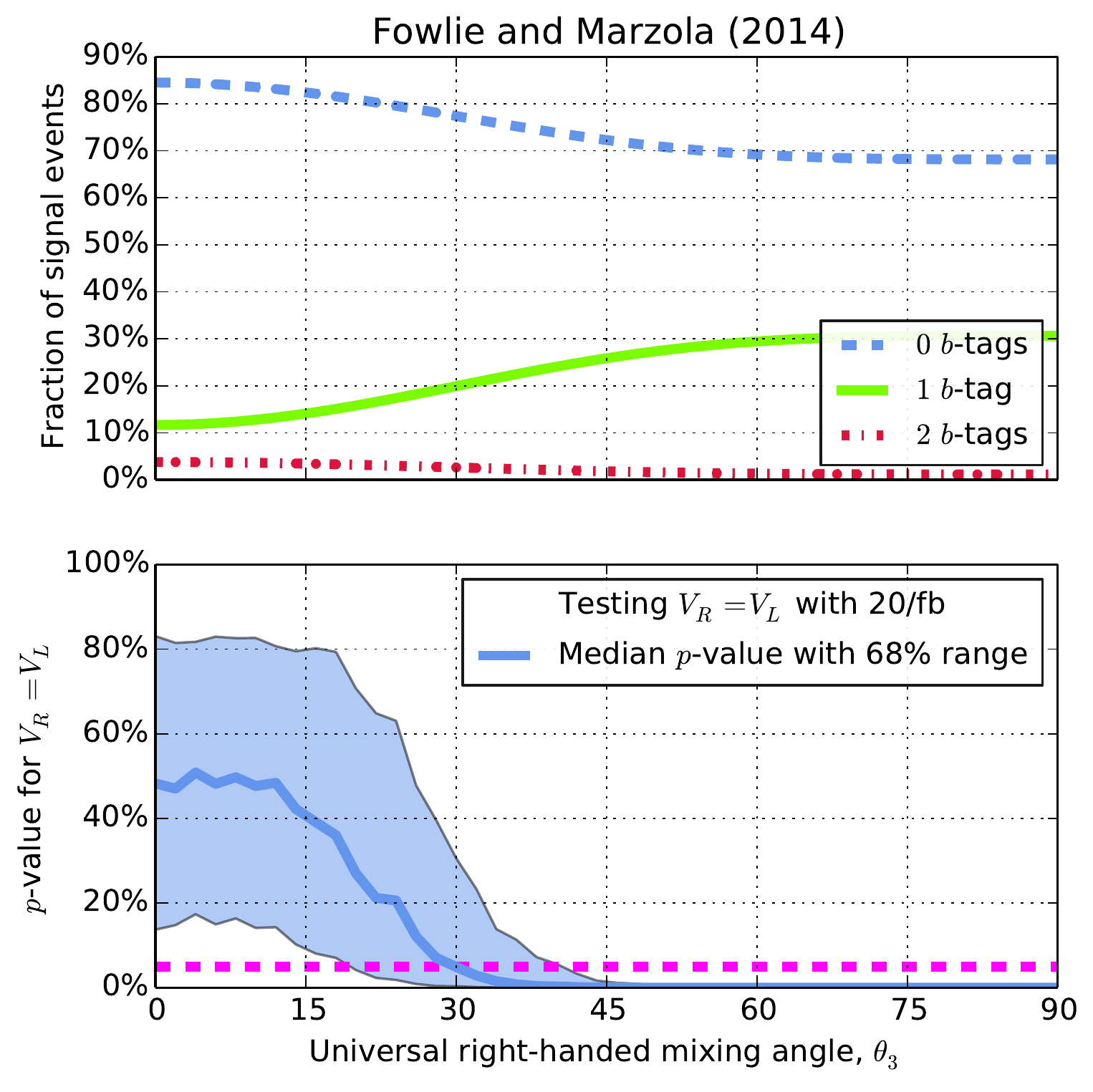}
\caption{Potential to exclude $V_R = V_L$, as a function of a universal mixing angle by counting $b$-tags at the LHC with \roots{13}, $\int\mathcal{L}\sim20\invfb$. When the $p$-value falls below $5\%$, we can reject the null hypothesis, that the RH quark mixing matrix is equal to the LH quark mixing matrix, with $95\%$ confidence. The upper panel shows the expected fractions of $b$-tags in the case in which $V_R\neq V_L$.
}
\label{fig:LR20}
\end{figure}

With $20\invfb$, we plot the $p$-value for the $V_R = V_L$ hypothesis against a universal mixing angle $\theta_3$ (with $\theta_4 = 0$) in \reffig{fig:LR20}.
The former can be rejected if $\theta_3 \gtrsim 30\degree$, confirming our preliminary analysis in FM1 which reported $\theta_3 \gtrsim 30\degree$. The behavior of the expected $b$-tag distribution (top panel in \reffig{fig:LR20}) as a function of the mixing angle is flatter than that in FM1, because of the imperfect efficiencies and purities achieved in our collider simulations, but the resulting exclusion is similar. In the upper panel in \reffig{fig:LR20}, we confirm that the $68\%$ range for the $p$-value shrinks rapidly, as found in our previous work.

\begin{figure}[t]
\centering
\includegraphics[width=.48\textwidth,valign=t]{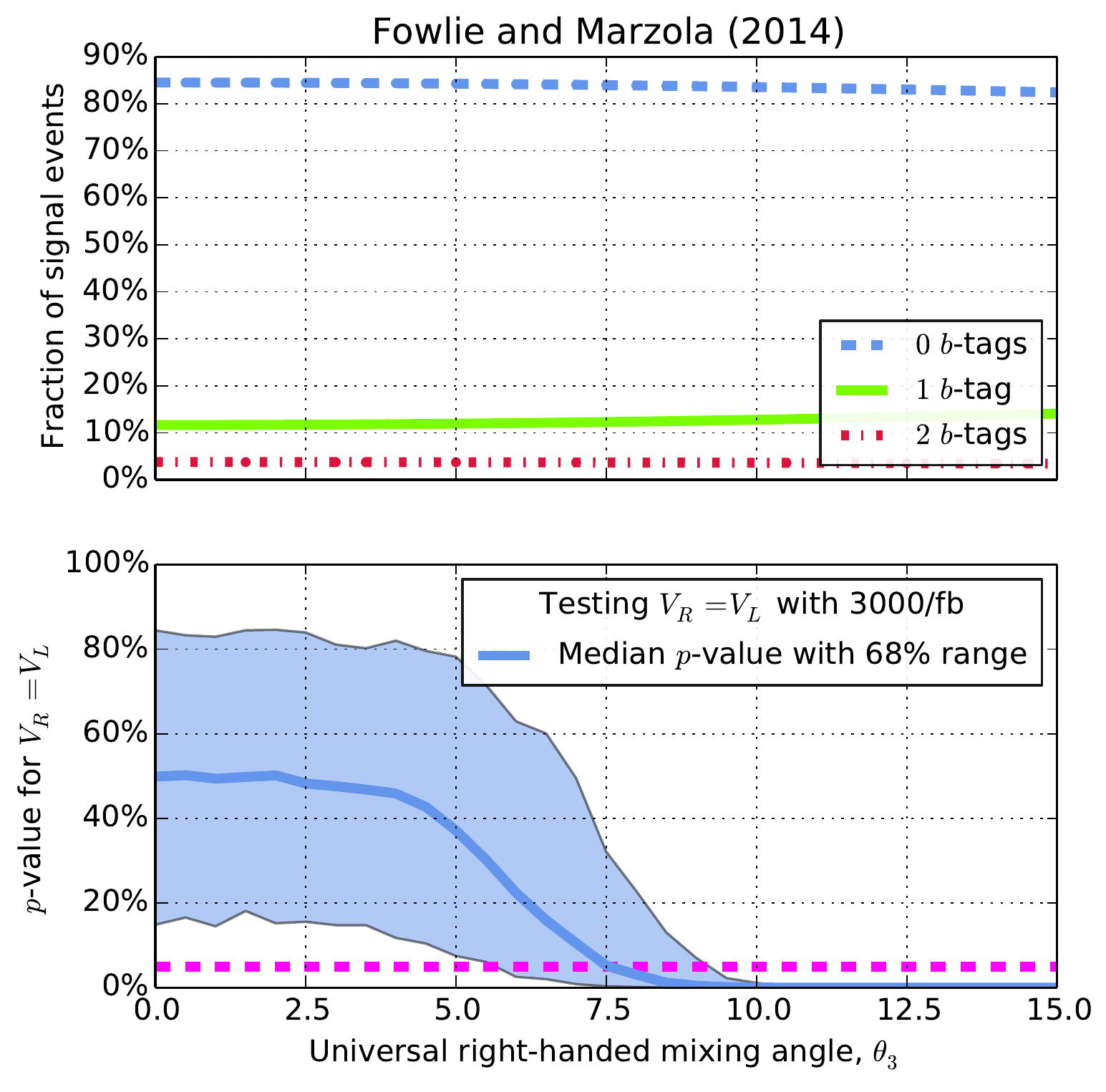}
\caption{As in \reffig{fig:LR20}, but with $3000\invfb$. Potential to exclude a unitary RH quark mixing matrix, $V_R = V_L$, as a function of a universal mixing angle by counting $b$-tags at the LHC with \roots{13}, $\int\mathcal{L}\sim3000\invfb$. When the $p$-value falls below $5\%$, we can reject the null hypothesis, that the RH quark mixing matrix is equal to the LH quark mixing matrix, with $95\%$ confidence. The upper panel shows the expected fractions of $b$-tags in the case in which $V_R\neq V_L$.
}
\label{fig:LR3000}
\end{figure}

In FM1, we also considered $3000\invfb$, and found that if the mixing angles were as small as about $7.5\degree$, our analysis could reject $V_R = V_L$. We repeat this test in \reffig{fig:LR3000}, confirming sensitivity to mixing angles as small as about $7.5\degree$. The differences between the crude analysis in FM1 and our refined analysis are negligible. As in the case with $20\invfb$, the $68\%$ interval for the $p$-value (filled blue band) shrinks rapidly.

\subsection{Possible improvements with a top-tagging algorithm}
Events involving top quarks can be identified by inspecting jet substructure and jet mass. This is known as top-tagging \see{Plehn:2011tg}. Top-tagging could improve the sensitivity of our analysis by expanding our categories with a $t$-tag category:
\beq
\label{Eq:4Cat}
\begin{pmatrix}
\text{Number $0$ $b$-tags} \\
\text{Number $1$ $b$-tags} \\
\text{Number $2$ $b$-tags} \\
\end{pmatrix} \xrightarrow{\text{top-tagger}}
\begin{pmatrix}
\text{Number $0$ $b$-tags and $0$ $t$-tags} \\
\text{Number $1$ $b$-tags and $0$ $t$-tags} \\
\text{Number $0$ $b$-tags and $1$ $t$-tags} \\
\text{Number $1$ $b$-tags and $1$ $t$-tags} \\
\end{pmatrix}
\eeq
An event is top-tagged if a combination of objects in that event satisfies a top-tagging algorithm; however, we forbid jets in such combinations from resulting in a further $b$-tag.

As well as extending our categories, a top-tagger may reduce the contamination between the former. For example, a $\rw\to tb$ event, which ought to be categorized as a two $b$-jet event, may contaminate the one $b$-tag category in our analysis, if, amongst other things, a secondary jet from the top-decay is selected. A top-tagger may reduce contamination in such events because, once a combination of objects is successfully top-tagged, there is no risk of wrongly selecting a jet from a $W$-boson decay resulting from the top decay, instead of the primary jet from the top decay.

We model the potential impact of a top-tagger by supposing that top-tagging algorithms increase the chances of correctly selecting the primary jet from a top-decay. We keep only three event categories, but assume that contamination between categories is reduced, which we implement with an improved efficiency matrix. We find that improvements in our efficiency matrix yield diminishing returns, at best extending our sensitivity to RH mixing angles by a few degrees.

We could, in principle, perform an analysis with a top-tagger and with the four categories in \refeq{Eq:4Cat}. The division of the one $b$-tag category into a $b$-tag category and a $t$-tag category might improve sensitivity. However, all RH mixing matrices that are symmetric modulo phases\footnote{If and only if $|V_{ij}| = |V_{ji}|$, then $V$ is symmetric modulo phases. Phases in the RH mixing matrix are irrelevant to our analysis.} (SMP) make approximately identical predictions for the ratio of the numbers of events in these two categories (the predictions are identical if there are no impurities). Thus, this extra information from the top-tagger might not help to discriminate between a unitary (and thus SMP) RH mixing matrix and an SMP non-unitary RH mixing matrix, as considered in our analysis.

\section{Conclusions}
Building upon our earlier work, we proposed methods with which one could examine a RH quark mixing matrix at the LHC at \roots{13}, if a \rw boson with a mass of about $2\tev$, hinted at by experiments at \roots{8}, were discovered in a final state with two electrons and two jets. Our methods involved counting the numbers of $b$-tags from a \rw boson decay. If the RH quark mixing matrix is unitary, particular relations between the branching fractions are required (other than the trivial condition that branching fractions sum to unity), which could, in principle, be checked.

With $20\invfb$, we find that the sensitivity of our test is somewhat limited by small statistics in the two $b$-tag category. One can exclude unitarity, but only for particular structures of RH quark mixing matrix; even maximal mixing with a fourth-generation quark by itself is insufficient. On the other hand, with $3000\invfb$, our method is able to significantly constrain departures from unitarity in the RH quark mixing matrix. If the anomaly observed in \refcite{Khachatryan:2014dka} persists and in the future a \rw boson is discovered, our test could be the first check of the unitarity of the RH quark mixing matrix. This would be an important test because rejecting unitarity would cast doubt upon the simplest interpretation of a \rw boson. We briefly discussed the application of this method to the unitarity of the LH CKM quark mixing matrix.

Concerning the possible inequality of the LH and RH quark mixing matrices, their difference is statistically significant with $20\invfb$ if the mixing angles in the RH quark mixing matrix are greater than about $30\degree$ and with $3000\invfb$, we achieve sensitivity to angles as small as about $7.5\degree$. Our findings with a full MC collider simulation are in agreement with our preliminary analysis FM1\cite{Fowlie:2014mza}. We suspect that our efficiencies and purities, especially for the two $b$-tag category, could be further optimized. If the anomaly persists, one challenge might be improving the efficiencies to maximize the insights into the RH quark mixing matrix.

\begin{acknowledgements}
We thank M.~Kadastik and S.~Moretti for their helpful comments and advice.
AF was supported in part by grants IUT23-6, CERN+, and by the European Union
through the European Regional Development Fund and by ERDF project 3.2.0304.11-0313
Estonian Scientific Computing Infrastructure (ETAIS).
LM acknowledges the European Social Fund for supporting his work with the grant MJD387.
\end{acknowledgements}

\bibliography{unitarity}
\end{document}